# Automated Artery Localization and Vessel Wall Segmentation of Magnetic Resonance Vessel Wall Images using Tracklet Refinement and Polar Conversion

Li Chen, *Student Member, IEEE*, Jie Sun, Gador Canton, Niranjan Balu, Xihai Zhao, Rui Li, Thomas S. Hatsukami, Jenq-Neng Hwang, *Fellow, IEEE*, Chun Yuan

*Abstract*— Quantitative analysis of vessel wall structures by automated vessel wall segmentation provides useful imaging biomarkers in evaluating atherosclerotic lesions and plaque progression time-efficiently. To quantify vessel wall features, drawing lumen and outer wall contours of the artery of interest is required. To alleviate manual labor in contour drawing, some computer-assisted tools exist, but manual preprocessing steps, such as region of interest identification and boundary initialization are needed. In addition, the prior knowledge of the ring shape of vessel wall is not taken into consideration in designing the segmentation method. In this work, trained on manual vessel wall contours, a fully automated artery localization and vessel wall segmentation system is proposed. A tracklet refinement algorithm is used to robustly identify the centerlines of arteries of interest from a neural network localization architecture. Image patches are extracted from the centerlines and segmented in a polar coordinate system to use 3D information and to overcome problems such as contour discontinuity and interference from neighboring vessels. From a carotid artery dataset with 116 subjects (3406 slices) and a popliteal artery dataset with 5 subjects (289 slices), the proposed system is shown to robustly identify the artery of interest and segment the vessel wall. The proposed system demonstrates better performance on the carotid dataset with a Dice similarity coefficient of 0.824, compared with traditional vessel wall segmentation methods, Dice of 0.576, and traditional convolutional neural network approaches, Dice of 0.747. This vessel wall segmentation system will facilitate research on atherosclerosis and assist radiologists in image review.

## I. INTRODUCTION

ATHEROSCLEROTIC cardiovascular disease is the leading cause of death worldwide [1]. Angiographic techniques are commonly used to depict luminal stenosis resulting from atherosclerosis progression. However, they often under- or over-estimate the underlying disease burden due to expansive or restrictive arterial wall remodeling that may be present [2]. Black-blood vessel wall magnetic resonance imaging (MRI) has allowed for direct visualization of atherosclerotic lesions in major arterial beds [3], [4] without ionizing radiation or contrast medium. Arterial wall segmentation in vessel wall MRI provides a quantitative analysis of atherosclerotic burden, which can be exploited for monitoring disease progression in serial studies and clinical trials [5], [6].

Most previous studies using quantitative analysis of vessel wall relied on manual vessel wall segmentation by drawing inner and outer boundaries of the arterial wall (lumen and outer wall) in each slice of MR images [7], which is tedious and subject to reader variability [8]. Identification of the region containing the artery of interest is usually needed before vessel wall segmentation; for example, zooming in to the region including the common carotid artery (CCA) or internal carotid artery (ICA) for carotid vessel wall analysis so that the thin vessel wall region can be better segmented. Semi-automated or automated methods have been proposed to segment vessel walls, such as using activate contour models by Yuan et al. [9] and Adams et al. [10], or using graph cut by Arias-Lorza et. al. [11], [12]. Another category of methods segment vessel wall area by classifying pixels into vessel wall regions and non-vessel wall regions using machine learning models [13], [14]. Manually locating the artery of interest is required for most methods, but some methods try to automatically locate arteries by referring to registered MR angiography, in which lumen areas are better visible [15]. In addition, Hough circle detection has been attempted to detect arterial centers under the

This work was supported in part by the National Institutes of Health under Grant R01 HL103609, American Heart Association under grant 18AIML34280043 and Philips Healthcare.

L. Chen and J. Hwang are from the Department of Electrical and Computer Engineering, University of Washington, Seattle, WA, 98195, USA (email: cluw@uw.edu, hwang@uw.edu).

J. Sun, G. Canton, N. Balu and C. Yuan are from the Department of Radiology, University of Washington, Seattle, WA, 98195, USA (email: sunjie@uw.edu, gcanton@uw.edu, ninja@uw.edu, cyuan@uw.edu).

X. Zhao, R. Li are from the Department of Biomedical Engineering, Tsinghua University School of Medicine, Beijing, China (email: xihaizhao@tsinghua.edu.cn, leerui@tsinghua.edu.cn).

T. Hatsukami is from the Department of Surgery, University of Washington, Seattle, WA, 98195, USA (email: tomhat@uw.edu).

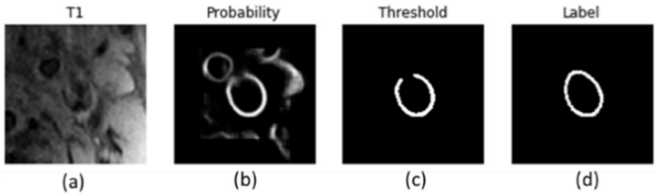

Fig. 1. Exemplar problems encountered previously in CAE [20]. (a) Original vessel wall image (b) Probability map from prediction. The external carotid artery is visible in the region of interest for internal carotid artery (the target artery) vessel wall segmentation, leading to both arteries with high probability. (c) Broken vessel wall segmentation due to weak signal on vessel wall region. (d) The human label.

assumption that arteries are circular in shape [16]. These methods reduce some manual steps and show reasonable agreement for images with high vessel-wall contrast. However, extensive human input is still needed for most methods, including seed-point/s initialization, artery of interest identification, and registration of image sequences. In addition, the robustness of the algorithm was not fully explored in previous studies likely due to the limited number of annotated samples in a specific vascular region.

Deep learning-based methods have shown superior performance in cardiovascular applications compared to traditional methods recently, including retinal blood vessel [17] and coronary artery segmentation [18]. In our previous work, the convolutional auto-encoder (CAE) structure demonstrated a high agreement with manual contours in lumen [19] and vessel wall [20] MRI segmentation. However, several major problems exist, preventing our deep learning-based algorithms from being effectively used: (1) The target artery cannot be automatically identified in the presence of multiple arteries. (2) Some prior knowledge, such as that vessel wall contours should be closed rings, is not used. Fig. 1. illustrates a problematic case. (3) Information from neighboring slices is not well used to refine the segmentation results.

In this study, we overcome the above challenges with a two-step fully automated vessel wall analysis workflow. A localization approach using tracklet refinement is developed to first robustly identify the lumen center of arteries along image slices to provide regions of interest for the subsequent vessel wall segmentation. Different from the commonly used Cartesian coordinate-based segmentation methods, we propose to transform the ring-shaped vessel wall to a polar coordinate system to ensure the continuity and accuracy of vessel wall boundaries.

In summary, the major contributions of this work are in four folds:

1) We propose a fully automated vessel wall segmentation workflow for black blood vessel wall MRI without any manual intervention. The use of an artery localization architecture to identify artery centerlines before vessel wall segmentation avoids the step of selecting the region of interests for arteries. The use of the convolutional neural network model for segmentation avoids the requirement for wall boundary initialization. At the same time, after vessel wall segmentation, the center location of the segmented lumen can be used to further refine the artery centerline, providing additional value for future analysis, such as arterial curved planar reformation.

2) The vessel wall segmentation performance in the polar coordinate system is extensively explored and compared with the Cartesian coordinate system. Different from almost all other methods based on the Cartesian coordinate system, the vessel wall segmentation in the polar coordinate system provides unique benefits, including better vessel wall continuity and improved segmentation especially needed in challenging slices near arterial bifurcations, where the artery shape is no longer circular.

3) The model takes advantage of 3D polar information along the arteries, showing improved performance than traditional 2D Cartesian segmentation methods. Neighboring slice information not only helps to construct a smooth and continuous artery centerline for better localization of the artery of interest, but also provides more information for more consistent vessel wall segmentation along slices, especially when a single slice of vessel wall is not clear enough for vessel wall segmentation.

4) The algorithm is developed based on a large dataset with more than one thousand subjects, and tested on both carotid and popliteal artery wall segmentation from different MR sequences. We have built and shared the popliteal vessel wall segmentation dataset, including the already publicly available MR images, segmentation results from the dataset, as well as manual segmentations from an experienced vessel wall reviewer. We welcome better segmentation methods to compare with our results, and encourage more research on vessel wall and atherosclerosis analysis using this dataset.

The rest of this paper is organized as follows. In Section II, we give a detailed description of the methodologies contained in our proposed localization and segmentation system. The experimental data setup and simulation results are given in Section III, followed by the discussions in Section IV. Finally, we draw the conclusion in Section V.

## II. PROPOSED LOCALIZATION AND SEGMENTATION METHODOLOGIES

The workflow for proposed localization and segmentation methodologies is shown in Fig. 2.

### A. Lumen Center Localization

The purpose of the localization task is to automatically identify the lumen center of each image slice to provide the region of interest for the subsequent vessel wall segmentation task. Tracking the artery across consecutive images is a similar task as object tracking in a video. Therefore, a tracking-by-detection approach is adopted in our localization scheme. For robust artery tracking, two steps, artery detection followed by tracklet refinement, are required.

For artery detection, a Yolo V2 detector [21] based on convolutional neural networks (CNN) is used to predict bounding boxes (minimum encompassing rectangles covering whole artery regions) of arteries in each image slice with an artery confidence score. The original weights of the Yolo detector are used to further train the model in artery detections.

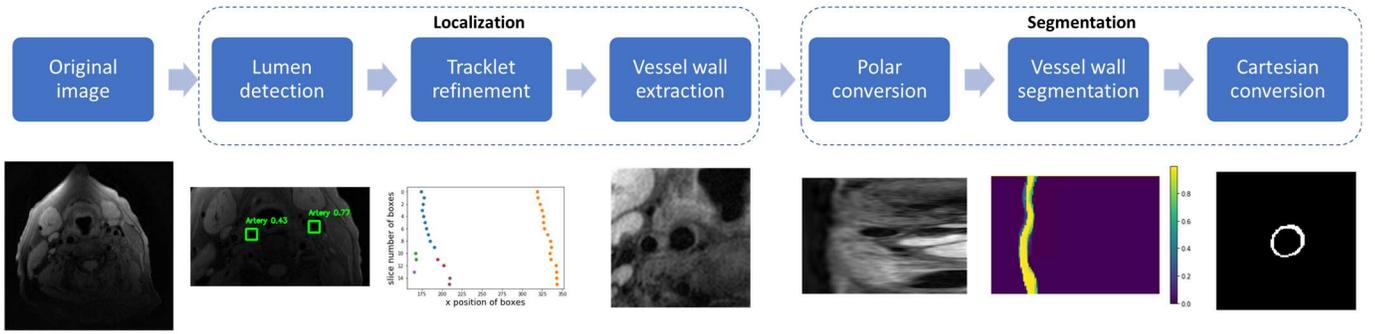

Fig. 2 Workflow for proposed localization and segmentation methodologies.

When no, or multiple bounding boxes are detected for some slices, a tracking method (tracklet refinement algorithm) is used to infer the missing detections or remove detections corresponding to arteries not of interest. It is assumed that arteries of interest will overlap in adjacent slices, which generally holds true for this application. Therefore, neighboring bounding boxes are grouped as tracklets if their Intersection of Union (IoU) is above a threshold. A connection loss for tracklet merging is defined as weighted sum of three losses, i.e., l1: loss for number of slices with missing detections between tracklets, l2: loss for non-overlapped region, and l3: loss for shape changes between bounding boxes. Tracklets are pairwise calculated for connection losses, and the pair with mutual minimum loss among all connection options are connected, with missing bounding boxes linearly interpolated. Artery confidence scores within the tracklet are summed up, and the tracklets with the top two scores are target tracklets. An example of using tracklet refinement to find missing detections and remove noise detections is shown in Fig. 3.

The centerline of the target tracklet is identified by connecting the center locations (i.e., 2D (x,y) position) along the slice (i.e., z position) of each bounding box within the tracklet. Constant image patch sizes of 128*128 are cropped along the centerline for subsequent vessel wall segmentation.

*B. A Polar Segmentation CNN Architecture*

A workflow for the polar segmentation system and the neural network structure are shown in Fig. 4., where an image patch along with the neighboring slices is linearly interpolated four times, the same image size as what human reviewers commonly use when drawing the contours. Polar transformation is used to convert coordinates of patches using the following functions:

$$\begin{cases} P[theta, rho] = I[y, x], \\ y = 256 + rho * \sin\left(theta * \frac{360}{PATCHHEIGHT}\right), \\ x = 256 + rho * \cos\left(theta * \frac{360}{PATCHHEIGHT}\right), \\ theta \in [0, PATCHHEIGHT = 180), \\ rho \in [0, PATCHWIDTH = 256). \end{cases} \quad (1)$$

A CNN architecture, similar to our previous CAE [20], is used to segment the vessel wall images in the polar system. Briefly, the network structure has three blocks of convolutional layers and max pooling layers, followed by three blocks of convolutional layers and up-sampling layers. Binary cross-entropy is used as the loss function. Adam optimizer [22] is used to control the learning rate.

Considering the similarity along the angle directions (sectors of vessel wall resemble each other), a sliding window to further crop smaller patches (height of 40) along the theta direction is used to reduce the network complexity and increase the samples for training. After predictions of all the patches, overlapped prediction regions are averaged as the probability map for the vessel wall. Initial contours of lumen and outer wall are defined as the contours from maximum and minimum gradient of the probability map in the radius direction. To identify smooth contours from the probability map, the active contour model (snake) [23] is applied. The region within lumen and outer wall contours is defined as vessel wall segmentation outcome, which is then converted back to the Cartesian coordinate system as the final segmentation result.

Segmentation confidence can be calculated from the prediction results to flag problematic slices in segmentation

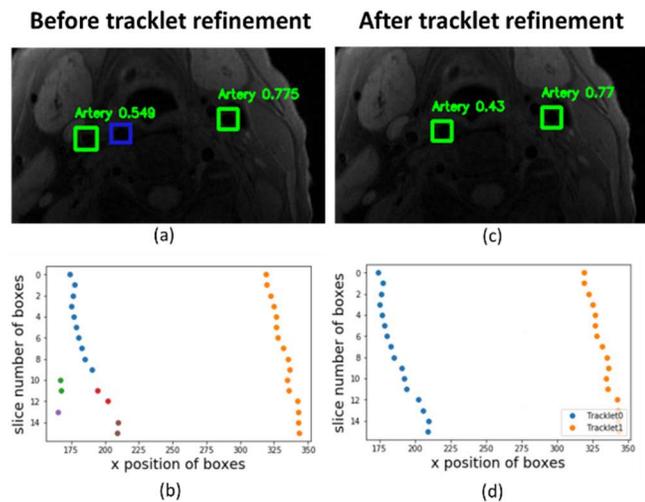

Fig. 3 (a) Before tracklet refinement, two targets are detected (in green boxes with artery confidence score), but one of them is wrong (ground truth in the blue box). (b) Detected centerline positions (only X position is shown) of bounding boxes along slices in tracklets. (c) Bounding boxes from both carotid arteries are found and wrong boxes are removed after tracklet refinement. (d) centerline positions of bounding boxes along slices in tracklets after refinement.

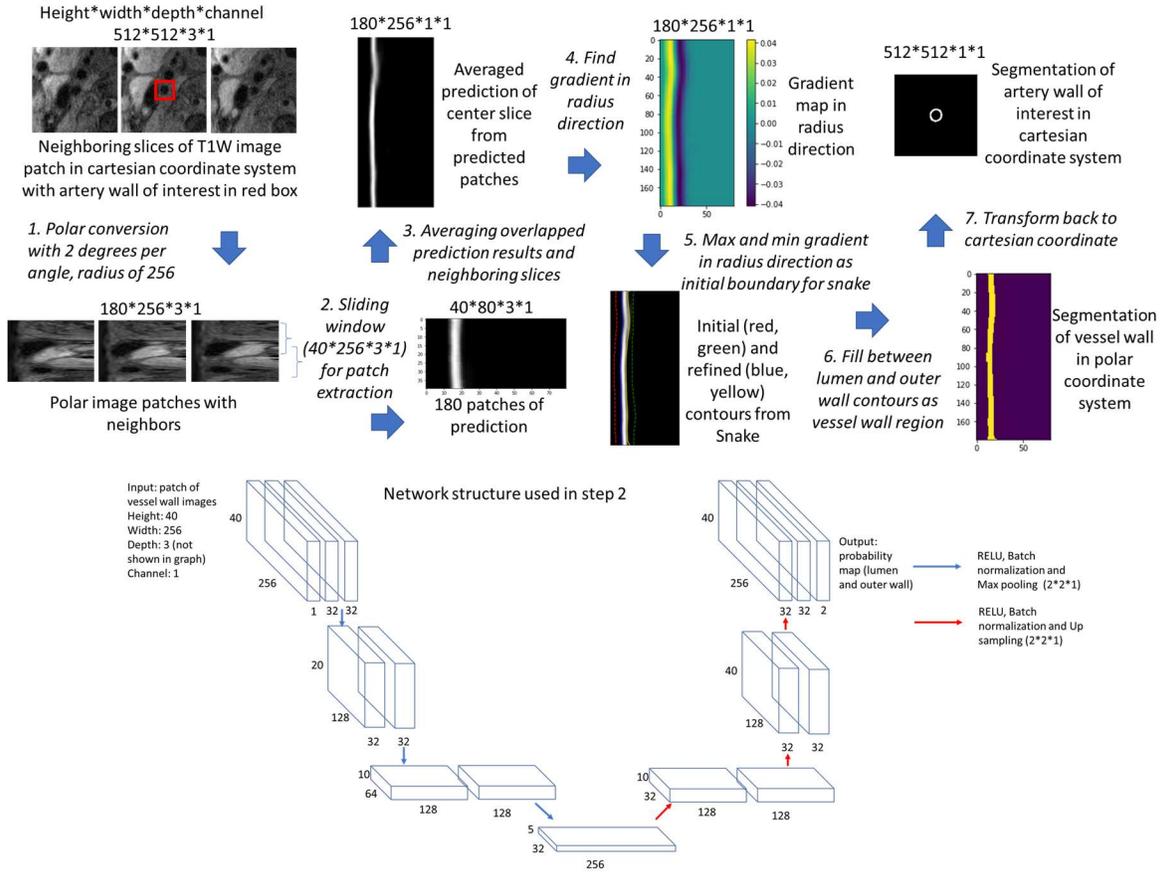

Fig. 4 Workflow for proposed polar segmentation CNN architecture.

where the model struggles in deciding boundaries. For each slice, the segmentation results are converted to a binary mask with 1 as segmented vessel wall, and -1 as the background. The segmentation confidence can be calculated using the following equation, by combining the binary mask $m_{i,j}$ and probability map $p_{i,j}$ where $i,j$ represent each pixel in the polar coordinate image.

$$Segconf = \frac{\sum p_{i,j} * m_{i,j}}{\sum m_{i,j} \,|m_{i,j} > 0}$$

(2)

Vessel wall segmentation with good agreement with manual contours demonstrates clear boundaries and a simple ring shape in MR images, and thus the segmentation neural network generates consistent vessel wall predictions from multiple overlapped patches, leading to sharp vessel wall boundaries in the probability map after aggregating patches. As a result, the predicted contours match exactly the boundaries in the probability map, and yield a segmentation confidence of 1. By checking slices with low confidence, readers can manually correct segmentations with uncertainty when needed.

### C. Centerline refinement from segmentation

The centerline generated from the localization model might not be in the real lumen center along slices. After the vessel wall segmentation, the center deviations can be calculated by comparing the segmented lumen boundaries in the polar coordinate system. The horizontal deviation is the difference in rho axis at angle 0 and 180 degrees, and at angle 90 and 270 degrees for the vertical direction.

By adjusting the centerline according to the deviations, another series of patches with lumen areas with better centers can be extracted for another round of segmentation. The process can be iteratively processed until the deviations in both directions are below the imaging resolution (4 pixels in this study).

TABLE 1
IMAGING PARAMETERS AND PROPERTIES FOR THREE DATASETS USED IN THIS STUDY

| | CARE-II | KOWA | OAI |
|---|---|---|---|
| Scanned artery region | Carotid artery | | Popliteal artery |
| Sequence | Fat suppressed T1 weighted Turbo Spin Echo | | 3D DESS |
| Blood suppression | Quadruple inversion recovery | | None |
| Number of subjects | 954 | 203 | 30 |
| Number of labeled slices | 27397 | 5194 | 1680 |
| TR (ms) | 800 | 800 | 16 |
| TE (ms) | 10 | 10 | 5 |
| In-plane Resolution (mm) | 0.57*0.57 | 0.63*0.63 | 0.36*0.36 |
| Spacing between slices (mm) | 2 | 2 | 1.5 |
| Field of view (mm*mm) | 160*160 | 160*160 | 140*140 |

## III. EXPERIMENTAL DATA SETUP AND RESULTS

### A. MR images

Data were collected following institutional review board guidelines. Informed consent was obtained from all study participants.

The first dataset includes T1-weighted (T1W) carotid artery images of 954 patients with recent ischemic stroke or transient ischemia attack, which were collected from the CARE-II study from multiple sites across China [24].

The second dataset includes T1W carotid artery images of 203 asymptomatic subjects from a clinical trial (NCT00851500; http://clinicaltrials.gov) for the Kowa Research Institute [25], [26].

The third dataset includes double echo steady state (DESS) popliteal artery images of 30 subjects from the Osteoarthritis Initiative (OAI) [27] sponsored by the National Institutes of Health, which is available for public access at http://www.oai.ucsf.edu/. The study is originally designed for knee osteoarthritis research, but the femoral-popliteal artery vessel wall in the MR scan is clearly visible, as discussed in [28].

Detailed imaging parameters are shown in TABLE 1.

### B. Human labels

Lumen and outer wall are traced manually by trained reviewers with >3 years' experience in cardiovascular MR imaging using a custom-designed software package (CASCADE) [29]. Image slices with poor image quality are excluded from review.

The CARE-II and Kowa datasets are pooled and randomly divided into 925 subjects (80%; 26008 image slices) as the training set, 116 subjects (10%; 3215 image slices) as the validation set, and 116 subjects (10%; 3406 image slices) as the testing set.

The OAI dataset is used to assess the generalizability of the algorithm on a different sequence and vascular bed. A training set including 23 subjects (1278 slices) is used to further tune the carotid system on popliteal arteries. The validation and test sets include 2 subjects (113 slices) and 5 subjects (289 slices), respectively.

Manual annotation time increases with increasing number of image slices. A human reviewer requires about one hour to annotate one subject of carotid artery scan, and 3 hours to annotate one subject with popliteal artery scan.

### C. Performance evaluation

The lumen center localization and vessel wall segmentation systems are trained and evaluated separately.

Mean IoU between detected and labeled bounding boxes, number of missing detections (labeled box has no overlap with any detections), and number of false detections (detected box does not have any overlap with any labels) are used to evaluate the performance of the localization system before and after tracklet refinement.

Performance of the segmentation is evaluated by the Dice

TABLE 2
VESSEL WALL SEGMENTATION PERFORMANCE COMPARED WITH TRADITIONAL METHODS

| Dataset | Model | DICE | CORRELATION | Time (s) |
|---|---|---|---|---|
| Carotid high quality | CASCADE[30] | 0.576 | N/A | N/A |
| | Polar CNN | 0.822 | N/A | N/A |
| Carotid | Cartesian CNN [20] | 0.747 | 0.870 | 0.03 |
| | Polar CNN | 0.824 | 0.885 | 40.33 |
| Popliteal | Cartesian CNN [20] | 0.695 | N/A | N/A |
| | Polar CNN | 0.821 | N/A | N/A |
| | Polar CNN with one slice | 0.808 | N/A | N/A |

similarity coefficient (DSC), which ranges from 0 (no overlap) to 1 (identical results). In addition, the correlation coefficient between manual and automated measurements of vessel wall area on each slice is calculated using the Pearson product-moment correlation method.

For fair comparison with the polar segmentation method, the Cartesian based segmentation with a similar CNN architecture is trained and tested using the same datasets. A traditional vessel wall segmentation algorithm implemented in CASCADE [30], which is based on automated B-spline snake and requires initiation by manually identified lumen centers, is also compared on subjects with highest level of image quality labeled during manual review from the carotid artery testing set. Note that the traditional B-spline method is shown to fail to detect vessel wall in low quality images.

To evaluate if the algorithm is generalizable, we apply this system on the OAI dataset with popliteal artery images. The same evaluation methods are used as in the carotid arteries.

To evaluate the contribution of neighboring slices to segmentation improvement, patches are repeated three times as the input to train the same segmentation neural network. The same testing set is used for evaluating the performance with each patch repeated three times as the input.

Model training and evaluation are performed on a workstation (Intel® Xeon® CPU E5-1650 v4 @3.6GHz 6 cores, 64 GB Memory) with NVIDIA Titan Xp GPU. Tensorflow [31] and Keras [32] are used as the deep learning platform in this study.

### D. Performance on test set

From validation sets, parameters are chosen to find the best performance for the tracklet refinement algorithm empirically. Before tracklet refinement, 94 bounding boxes from manual labels are not detected, and an additional 32 bounding boxes are detected for other than the artery of interest. The mean IoU between detected and labeled bounding boxes is 0.779. After tracklet refinement the mean IoU is 0.791, with 68 missed detections, and all the wrongly detected boxes are removed.

The mean DSCs for polar and Cartesian segmentation are 0.824 and 0.747. The correlation coefficient of vessel wall area between polar segmentation results and the ground truth is 0.885. As a comparison, the correlation coefficient for the

Cartesian segmentation is 0.870. As an example, the segmentation results by the two methods on two image slices near carotid bifurcation are shown in Fig. 5.

Among all carotid arteries in the testing set, 7 image scans (111 slices in total) have the highest level of image quality for all the slices. The mean DSC for the traditional segmentation method is 0.576, excluding 5 slices of failed segmentations. The mean DSC from the polar segmentation CNN architecture is 0.822 with no failures.

The system is able to identify popliteal artery locations and segment artery walls with only a small training set when transferred from the carotid arteries. The mean IoU for bounding boxes is 0.873 and the mean DSC is 0.821. The DSC decreases to 0.808 if only a single slice is used as the input. Mean segmentation time for each slice is 40.3 seconds, with most time used in coordinate transformation.

The comparison of segmentation models and their performance are summarized in TABLE 2.

## IV. DISCUSSION

In this study, fully automated vessel wall segmentation has been achieved with a high accuracy by effectively using CNN models to detect artery locations using tracking-by-detection approach and segment vessel wall in the polar coordinate system. The step of artery localization avoids the manual procedure to select the region for artery analysis so that the vessel wall can be analyzed with no human intervention. Traditional vessel wall segmentation methods are susceptible to poor image quality, only providing reasonable results when both lumen and outer wall boundaries have high contrast. Our deep learning-based method extracts useful boundary information from more than 32,000 slices of manually drawn vessel wall contours with various levels of image quality. We believe our dataset has encompassed a wide spectrum of atherosclerosis to train a robust deep learning model with good generalizability. The development of polar segmentation CNN architecture, adding more prior knowledge of vessel wall structures (e.g. ring shape, lumen center), takes advantage of both human defined and machine learned features and outperforms our previous deep-learning segmentation method based on a Cartesian coordinate system.

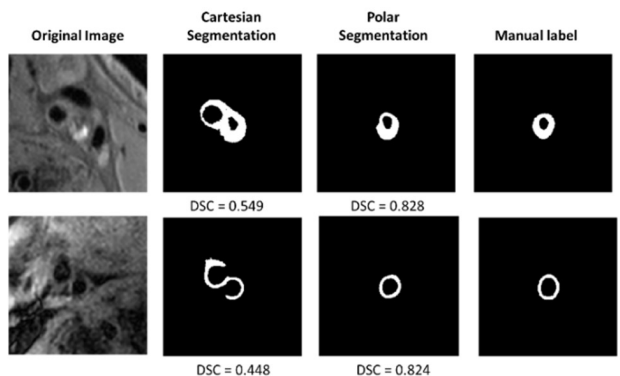

Fig. 5. Image slices near carotid bifurcations with (first row) and without (second row) plaque. Cartesian segmentation leads to wrong artery walls being segmented and cannot ensure wall continuity. The proposed polar segmentations have better agreement with human labels.

The application of deep learning methods in vessel wall segmentation might have a profound impact on MR vessel wall image analysis. As a research tool, with accurately segmented vessel wall areas from the automated method, quantitative vessel wall features can be extracted to enhance our understanding of atherosclerosis progression from large population studies, for which time-consuming manual or semi-automated methods are not achievable. Clinically, a fast screening tool can be developed to automatically flag high-risk patients for further detailed examination in a timely efficient manner.

The extra calculation time mainly for polar conversion and restoration is a major drawback for the current segmentation system. GPU acceleration for polar transformation can be attempted in the future. Additionally, only relatively straight carotid and popliteal arteries are evaluated in this study. However, the method has the potential to be adapted to MRI data of more tortuous arteries (e.g. intracranial arteries) with the combination of robust artery tracing and cross-sectional slicing methods [33].

## V. CONCLUSION

A deep learning system for vessel wall localization and segmentation has been developed with tracklet refinement and polar transformation. Compared with traditional methods, the proposed system avoids human intervention and demonstrates better performance in accurate segmentation of vessel wall areas. It has the potential to facilitate research on atherosclerosis and assist radiologists in image review.

### ACKNOWLEDGMENT

We are grateful to the support of NVIDIA Corporation with donation of GPU. We acknowledge the CARE-II, Kowa, and OAI researchers for the imaging data.